\documentclass[preprint,3p,times]{elsarticle}

\usepackage{graphicx}
\usepackage{multirow}
\usepackage{amsmath,amssymb,amsfonts}
\usepackage{amsthm}
\usepackage{mathrsfs}
\usepackage{hyperref}
\usepackage[title]{appendix}
\usepackage{xcolor}
\usepackage{textcomp}
\usepackage{manyfoot}
\usepackage{booktabs}
\usepackage{algorithm}
\usepackage{algorithmicx}
\usepackage{algpseudocode}
\usepackage{listings}
\usepackage{float}
\usepackage{url}

\newcommand{\R}{\mathbb{R}}
\newcommand{\mrm}[1]{\mathrm{#1}}
\renewcommand{\d}[1]{\:\mathrm{d}#1}
\newcommand{\T}{\mathrm{T}}
\newcommand{\assembly}{\mathop{\vphantom{\sum}\mathchoice{\vcenter{\hbox{\huge A}}}{\vcenter{\hbox{\Large A}}}{\mathrm{A}}{\mathrm{A}}}\displaylimits}

\newcommand{\C}{\mathbb{C}}
\newcommand{\bSigma}{\boldsymbol{\Sigma}}
\newcommand{\bsigma}{\boldsymbol{\sigma}}
\newcommand{\bEpsilon}{\boldsymbol{E}}
\newcommand{\bepsilon}{\boldsymbol{\varepsilon}}
\newcommand{\bu}{\boldsymbol{u}}
\renewcommand{\bf}{\boldsymbol{f}}
\newcommand{\bK}{\boldsymbol{K}}
\newcommand{\brho}{\boldsymbol{\rho}}
\newcommand{\bx}{\boldsymbol{x}}
\newcommand{\bgamma}{\boldsymbol{\gamma}}
\newcommand{\bkappa}{\boldsymbol{\kappa}}
\newcommand{\bG}{\boldsymbol{G}}
\newcommand{\bp}{\boldsymbol{p}}
\newcommand{\bW}{\boldsymbol{W}}
\newcommand{\bv}{\boldsymbol{v}}
\newcommand{\bb}{\boldsymbol{b}}

\newcommand{\rhomin}{\rho_{\mathrm{min}}}
\newcommand{\rhomax}{\rho_{\mathrm{max}}}
\newcommand{\rhovoid}{\rho_{\mathrm{void}}}

\newcommand{\nel}{n_{\mathrm{el}}}

\journal{Computer Methods in Applied Mechanics and Engineering}

\begin{document}

\begin{frontmatter}



\title{Multiscale topology optimization of functionally graded lattice structures based on physics-augmented neural network material models}

\author[a]{Jonathan Stollberg\corref{cor}}\ead{jonathan.stollberg@tu-darmstadt.de}
\author[b]{Tarun Gangwar}\ead{tarun.gangwar@ce.iitr.ac.in}
\author[c]{Oliver Weeger}\ead{weeger@cps.tu-darmstadt.de}
\author[a]{Dominik Schillinger}\ead{dominik.schillinger@tu-darmstadt.de}
\cortext[cor]{Corresponding author.}

\affiliation[a]{
	organization={Institute for Mechanics, Computational Mechanics Group, Technical University of Darmstadt},
	city={Darmstadt},
	postcode={64287}, 
	country={Germany}}          
\affiliation[b]{
	organization={Department of Civil Engineering, Indian Institute of Technology Roorkee},
	city={Roorkee},
	postcode={247667},
	state={Uttarakhand}, 
	country={India}}
\affiliation[c]{
	organization={Cyber-Physical Simulation, Technical University of Darmstadt},
	city={Darmstadt},
	postcode={64293}, 
	country={Germany}}  

\begin{abstract}
We present a new framework for the simultaneous optimiziation of both the topology as well as the relative density grading of cellular structures and materials, also known as lattices. Due to manufacturing constraints, the optimization problem falls into the class of NP-complete mixed-integer nonlinear programming problems. To tackle this difficulty, we obtain a relaxed problem from a multiplicative split of the relative density and a penalization approach. The sensitivities of the objective function are derived such that any gradient-based solver might be applied for the iterative update of the design variables. In a next step, we introduce a material model that is parametric in the design variables of interest and suitable to describe the isotropic deformation behavior of quasi-stochastic lattices. For that, we derive and implement further physical constraints and enhance a physics-augmented neural network from the literature that was formulated initially for rhombic materials. Finally, to illustrate the applicability of the method, we incorporate the material model into our computational framework and exemplary optimize two-and three-dimensional benchmark structures as well as a complex aircraft component.
\end{abstract}




\end{frontmatter}


\section{Introduction}
\label{sec:introduction}
Recent advances in manufacturing, such as 3D printing -- commonly referred to as additive manufacturing -- offer the possibility to overcome the limitations of traditional manufacturing techniques, thus allowing the creation of complex topology structures by adding slices of solid material layer by layer \citep{Gibson_2015}. Consequently, cellular materials such as functionally graded lattices have come to the forefront of both academia and industry due to their unique and superior properties in terms of stiffness-to-weight ratio, thermal conduction, and energy absorption \citep{Kim_2004, Grossmann_2019, Zhou_2019, Niknam_2020, Alkebsi_2021}. This trend is accompanied by the development of new topology and material optimization methods to fully exploit the advantageous properties of cellular lightweight structures in an application-specific manner as outlined by Zhu et al. \citep{Zhu_2021}.

The topology optimization of cellular multiscale structures as a design tool was pioneered by Bends{\o}e and Kikuchi \cite{Bendsoe_1988} with the development of the homogenization method. The idea of implementing a power-law as a material model, depending on a density variable into a topology optimization framework, came shortly after with the well-established Solid Isotropic Material with Penalization (SIMP) method \citep{Bendsoe_1989}. SIMP, however, is primarily used to compute void-solid designs, known as 0-1 designs, since porous structures resulting from this approach with a linear density-stiffness relation violate the Hashin-Shtrikman bounds \citep{Bendsoe_1999}. A less restricted solution space for the design of cellular structures is obtained through concurrent approaches \citep{Liu_2008, Xia_2014, Garner_2019}. In these approaches, one additional topology optimization problem per macroscale material point is solved on the microscale, resulting in complex unit cell layouts. Nevertheless, this requires extensive computational resources due to the heavily increased number of design variables \citep{Coelho_2008}. In contrast, homogenization-based topology optimization involving lattice unit cells of predefined geometry, which are parametric only with respect to the relative density, reduces the number of design variables to a minimum. This makes such methods computationally attractive, and therefore they constitute still an active field of research \citep{Wang_2017, Li_2018, Alkhatib_2023, Stroemberg_2024}.

Data-driven methods, such as artificial neural networks (ANNs), form a relatively new approach to solving engineering problems and are gaining more attention in the computational mechanics community \citep{Kollmannsberger_2021}. Generally, ANNs are highly flexible functions that map a set of input parameters to some output, making them particularly suitable for material modeling. For the first time this was implemented by Ghaboussi et al. \citep{Ghaboussi_1991} using a black-box approach without taking any physical conditions into account. Later, Shen et al. \citep{Shen_2004} used isotropic invariants as model inputs to incorporate basic constitutive constraints such as thermodynamic consistency and objectivity. These ANN models, that guarantee physically sensible outputs, are commonly known as physics-informed neural networks (PINNs) \citep{Raissi_2019, Bastek_2023}, physics-augmented neural networks (PANNs) \citep{Klein_2022, Rosenkranz_2024} or constitutive artificial neural networks (CANNs) \citep{Linka_2021, Linka_2023} in the literature. Over time, researchers have proposed many network architectures that fulfill different physical conditions of elastic \citep{Linden_2023} and inelastic material behavior \citep{Rosenkranz_2023}. This trend, furthermore, has led to the development of neural networks modeling the effective behavior of (parametric) lattice materials \citep{Gaertner_2021, Shojaee_2023} as alternatives to conventional linear homogenization approaches \citep{Ashby_2006, Vigliotti_2012, Huber_2018}. However, even though PANNs have already been applied to predict lattice architectures with specific target stiffness properties \citep{Bastek_2022}, only physically unconstrained ANN material models have been used for density-based topology optimization of cellular structures and materials to the best of our knowledge \citep{White_2019, Liu_2023, Kim_2024}.

In this article, we develop an optimization framework that simultaneously optimizes the topology and the material of hierarchical cellular structures and involves PANN material models. We structure the article as follows: In Section~\ref{sec:topology_optimization}, the general form of our topology optimization method is presented, including the derivation of the sensitivities of the objective function with respect to the design variables and a heuristic design variable update scheme that is simple to implement. Next, we discuss the data-driven material model in Section~\ref{sec:material_modeling}, focusing on the derivation of physical constraints and the generation of training data from computational homogenization. In Section~\ref{sec:results}, we include the material model in the optimization problem and compute two- and three-dimensional example structures to examine the applicability of our method. Finally, we draw conclusions from our results in Section~\ref{sec:conclusion}, along with an outlook on possible future research avenues.

\section{Topology optimization of functionally graded cellular structures}
\label{sec:topology_optimization}
In this section, we introduce a minimum compliance optimization problem along with a corresponding solution algorithm designed to optimize both the topology and material grading of cellular structures. Although our discussion primarily focuses on strut-based lattices, the algorithm is broadly applicable to various types of cellular materials, including gyroid lattices and metal foams.

\subsection{Design variables}
\label{sec:design_variables}
According to Ashby \citep{Ashby_2006}, the properties of cellular structures and materials depend on three groups of design variables, namely
\begin{enumerate}
	\item the topology and shape of the unit cell,
	\item the solid base material, which the cellular material is made of, and
	\item the relative density.
\end{enumerate}

\begin{figure}[htb]
	\centering
	\includegraphics[scale=0.95]{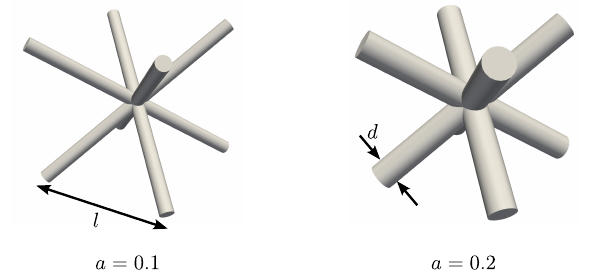}
	\caption{Strut-based body-centered cubic (BCC) unit cells for two different aspect ratios $a$. The volume filled with base material in the unit cell increases with the aspect ratio, implying the existence of a mapping function between the relative density and the aspect ratio, and vice versa.}
	\label{fig:bcc_cells}
\end{figure}

In our study, we will focus on the relative density as a design variable. The relative density is defined as the ratio of the density of the cellular material $\rho_{\mrm{cell}}$ over the density of the base material $\rho_{\mrm{base}}$ and describes the fraction of the unit cell volume that is actually filled with base material, i.e.
\begin{align}
	\label{eq:relative_density}
	\rho = \frac{\rho_{\mrm{cell}}}{\rho_{\mrm{base}}} = \frac{V_{\mrm{base}}}{V_{\mrm{cell}}}.
\end{align}
When referring to strut-based lattices, the relative density is related directly to the aspect ratio
\begin{align}
	\label{eq:aspect_ratio}
	a = \frac{d}{l}
\end{align}
such that in general $\rho = \rho \left( a \right)$ or equivalently $a = a \left( \rho \right)$ \citep{Cui_2018, Zhang_2023}. In~\eqref{eq:aspect_ratio}, $d$ denotes the strut diameter and $l$ is the unit cell size. In Figure~\ref{fig:bcc_cells}, this is exemplified using body-centered cubic (BCC) unit cells with different aspect ratios.

From~\eqref{eq:aspect_ratio}, we can conclude that in case the unit cell topology and strut diameter are uniform across the domain of the structure, grading with respect to the relative density also yields grading with respect to $l$. Consequently, a constant strut length or unit cell size yields grading with respect to $d$.

\subsection{General problem formulation}
\label{sec:problem_formulation}
Consider a cellular body $\Omega$ with Dirichlet boundary $\Gamma_{\mrm{D}}$ and Neumann boundary $\Gamma_{\mrm{N}}$ that is discretized with $\nel$ finite elements $e$ such that $\Omega_{e} \subset \Omega$ and
\begin{align}
	\label{eq:discretization}
	\bigcup_{e = 1}^{\nel} \Omega_{e} \approx \Omega.
\end{align} 
Each element is assigned a constant relative density $\rho_{e}$, such that a global density vector
\begin{align}
\label{eq:density_vector}
	\brho = \begin{bmatrix}
		\rho_{1} & \rho_{2} & \dots & \rho_{\nel - 1} & \rho_{\nel}
	\end{bmatrix}^{\T}
\end{align}
can be defined. The discrete form of the general minimum compliance topology optimization problem then reads \citep{Bendsoe_2004}
\begin{subequations}
\label{eq:general_optimization_problem}
\begin{align}
  		\min_{\brho}  &\hspace{0.5em} c\left(\brho\right) = \bf^{\T} \bu \label{eq:objective_function}\\
  		\text{s.t.}   
  		&\hspace{0.5em} \bK \bu = \bf, \label{eq:constraint_kinematic_admissibility}\\
  		& \hspace{0.5em} \frac{1}{\lvert \Omega \rvert} \sum_{e=1}^{\nel} \rho_{e} \lvert \Omega_{e} \rvert - V_{\mrm{frac}} \leq 0, \label{eq:constraint_volume}\\
  		& \hspace{0.5em} \rho_{e} \in \mathcal{A}_{i}, \quad \forall e \in \{ 1,\dots,\nel \}. \label{eq:constraint_admissible_values}
\end{align}
\end{subequations}
The equality constraint~\eqref{eq:constraint_kinematic_admissibility} serves the kinematic admissibility of the displacement vector $\boldsymbol{u}$ and follows directly from finite element analysis:
\begin{align}
\label{eq:stiffness_matrix}
	\bK = \assembly_{e=1}^{\nel} \int_{\Omega_{e}} \boldsymbol{B}^{\T} \, \overline{\C} \, \boldsymbol{B} \d{\Omega_{e}}
\end{align}
is the stiffness matrix obtained from an assembly operation and element-wise integration of shape function gradients $\boldsymbol{B}$ and the constitutive material tensor $\overline{\C}$, where the overline denotes compatibility with Voigt notation. The global vector $\bf$ denotes the external forces acting on $\Omega$. In addition, the fraction of the volume in the design space that is actually filled with solid material is limited to $V_{\mrm{frac}}$ through the inequality constraint~\eqref{eq:constraint_volume}, where $\lvert \Omega_{e} \rvert$ represents the volume of element $e$ and $\lvert \Omega \rvert$ the total volume of $\Omega$.

\begin{figure}[htb]
	\centering
	\includegraphics[scale=0.95]{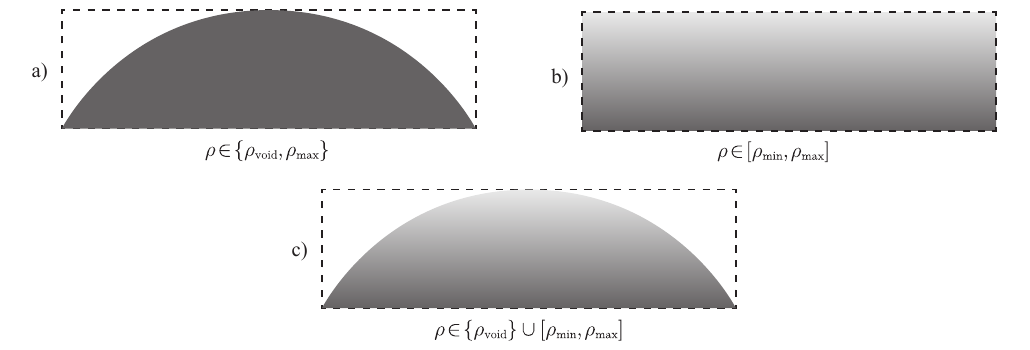}
	\caption{Admissible sets of relative density values and their effect on the optimized design. a) Simple 0-1 design. b) Design with graded density. c) Design with optimized topology and density grading.}
	\label{fig:optimization_approaches}
\end{figure}

The admissible relative density values are specified in~\eqref{eq:constraint_admissible_values} through the set $\mathcal{A}_{i}$, whose choice depends on the type of application or on the manufacturing constraints.  More specifically, to obtain simple 0-1 designs, the choice $\mathcal{A}_{1} = \left\{ \rhovoid, \rhomax \right\}$ is sufficient with $\rhovoid$ representing void, i.e., the absence of any solid or cellular material. To maintain numerical stability, $\rhovoid$ may not be exactly but close to zero. In this work, we choose $\rhovoid = 10^{-15}$. On the contrary, the set $\mathcal{A}_{2} = \left[ \rhomin, \rhomax \right]$ is suitable to optimize the grading of a cellular material with respect to the density in the given range as sketched in Figure~\ref{fig:optimization_approaches}. However, in order to optimize both the macroscopic topology of $\Omega$ as well as the material grading simultaneously, an admissible set
\begin{align}
	\label{eq:admissible_set}
	\mathcal{A}_{3} = \left\{ \rhovoid \right\} \cup \left[\rhomin , \rhomax \right]
\end{align}
with $0 < \rhovoid \ll \rhomin < \rhomax \leq 1$ might be used \citep{Lu_2023}. The continuous range $\left[\rhomin , \rhomax \right]$ covers all relative density values that are assumed to be manufacturable.

\subsection{Relaxation of the initial mixed-integer nonlinear programming problem}
\label{sec:relaxation}
For $\mathcal{A}_{i} = \mathcal{A}_{3}$, the design variables $\rho_{e}$ in the general problem become semi-continuous and, as a consequence,~\eqref{eq:general_optimization_problem} constitutes a mixed-integer nonlinear programming (MINLP) problem. MINLP problems are characterized as NP-complete in terms of complexity and require specialized algorithms to obtain a minimum \citep{Floudas_1995}. Our approach to address this difficulty is to split the relative density into a topological variable $\gamma$ and another density variable $\kappa$, which we refer to as microstructural relative density in the following. Hence, the relative density of finite element $e$ is 
\begin{align}
	\label{eq:relative_density_split}
	\rho_{e} = \gamma_{e} \kappa_{e}.
\end{align}
Analogously to~\eqref{eq:density_vector}, the newly introduced variables are stored in vectors $\bgamma$ and $\bkappa$ such that a relaxed optimization problem can be constructed:
\begin{subequations}
\label{eq:relaxed_problem}
\begin{align}
  		\min_{\bgamma, \bkappa}  &\hspace{0.5em} c\left( \bgamma, \bkappa \right) = \bf^{\T} \bu \label{eq:relaxed_objective_function}\\
  		\text{s.t.}   
  		&\hspace{0.5em} \bK \bu = \bf, \label{eq:relaxed_constraint_kinematic_admissibility}\\
  		& \hspace{0.5em} \frac{1}{\lvert \Omega \rvert} \sum_{e=1}^{\nel} \gamma_{e} \kappa_{e} \lvert \Omega_{e} \rvert - V_{\mrm{frac}} \leq 0, \label{eq:relaxed_constraint_volume}\\
  		& \hspace{0.5em} \gamma_{e} \in \left[ \rhovoid, 1 \right], \quad \forall e \in \{ 1,\dots,\nel \}, \label{relaxed_constraint_admissible_values_gamma}\\
  		& \hspace{0.5em} \kappa_{e} \in \left[ \rhomin, \rhomax \right], \quad \forall e \in \{ 1,\dots,\nel \}. \label{relaxed_constraint_admissible_values_kappa}
\end{align}
\end{subequations}

\begin{figure}[htb]
	\centering
	\includegraphics[scale=0.95]{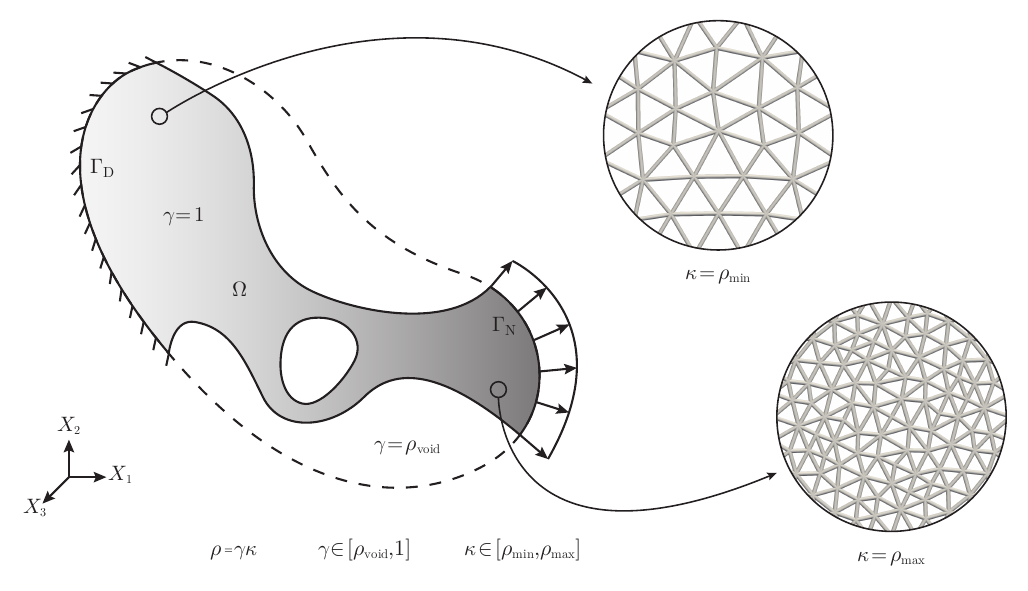}
	\caption{Multiplicative split of the relative density for the optimization of the topology and material grading of a lattice structure.}
	\label{fig:optimization_potatoe}
\end{figure}

To push $\gamma_{e}$ towards its limit values, a penalization approach similar to the well-established SIMP method is utilized \citep{Bendsoe_1989}. For that, the constitutive material tensor is defined element-wise as
\begin{align}
	\label{eq:penalized_material_tensor}
	\overline{\C} \left( \gamma_{e}, \kappa_{e} \right) = \gamma_{e}^{p} \; \overline{\C}^{*} \! \left( \bp \left( \kappa_{e}, \dots \right) \right)
\end{align}
with penalty parameter $p > 1$ and the effective stiffness tensor $\overline{\C}^{*}$ as an implicit function of the microstructural relative density and possibly further material parameters that set up input vector $\bp$. Figure~\ref{fig:optimization_potatoe} illustrates the role of the alternative set of design variables in the context of topology optimization.

We emphasize that in contrast to the initial MINLP problem, the relaxed problem~\eqref{eq:relaxed_problem} becomes solvable efficiently by means of standard gradient-based methods such as the method of moving asymptotes (MMA) \citep{Svanberg_1987, Svanberg_2002} or the optimality criteria method \citep{Bendsoe_2004}.

\subsection{Sensitivity analysis and stabilization}
\label{sec:sensitivity_analysis}
The sensitivity or gradient of the objective function $c$ with respect to the design variables $\bgamma$ and $\bkappa$ can be computed effectively by means of the adjoint method as demonstrated by Bends{\o}e and Sigmund \citep{Bendsoe_2004}:
\begin{subequations}
\label{eq:gradient_adjoint}
\begin{align}
	\frac{\partial c}{\partial \gamma_{e}} &= -\bu_{e}^{\T} \, \frac{\partial \bK_{\! e}}{\partial \gamma_{e}} \, \bu_{e}, \label{eq:gradient_gamma}\\
	\frac{\partial c}{\partial \kappa_{e}} &= -\bu_{e}^{\T} \, \frac{\partial \bK_{\! e}}{\partial \kappa_{e}} \, \bu_{e}. \label{eq:gradient_kappa}
\end{align}
\end{subequations}
The element displacement vector is denoted $\bu_{e}$ and the element stiffness matrix $\bK_{\! e}$ results from the integral in~\eqref{eq:stiffness_matrix}. Therefore, we rewrite~\eqref{eq:gradient_adjoint} as
\begin{subequations}
\label{eq:gradient_detailed}
\begin{align}
	\frac{\partial c}{\partial \gamma_{e}} &= -\bu_{e}^{\T} \left[ \int_{\Omega_{e}} p \, \gamma_{e}^{p-1} \, \boldsymbol{B}^{\T} \, \overline{\C}^{*} \, \boldsymbol{B} \d{\Omega_{e}} \right] \bu_{e}, \label{eq:gradient_gamma_detailed}\\
	\frac{\partial c}{\partial \kappa_{e}} &= -\bu_{e}^{\T} \left[ \int_{\Omega_{e}} \gamma_{e}^{p} \, \boldsymbol{B}^{\T} \, \frac{\partial \, \overline{\C}^{*}}{\partial \kappa_{e}} \, \boldsymbol{B} \d{\Omega_{e}} \right] \bu_{e}. \label{eq:gradient_kappa_detailed}
\end{align}
\end{subequations}
In order to keep the formulation of the effective stiffness tensor $\overline{\C}^{*}$ flexible, we compute the right-hand side partial derivative in~\eqref{eq:gradient_kappa_detailed} using efficient forward-mode automatic differentiation. 
The sensitivity numbers are then obtained from
\begin{align}
\label{eq:sensitivity_numbers}
	\alpha_{e}^{\gamma} = -\frac{\partial c}{\partial \gamma_{e}}, \quad \alpha_{e}^{\kappa} = -\frac{\partial c}{\partial \kappa_{e}}.
\end{align}

Topology optimization suffers from mesh-dependency and checkerboard patterns which can be tackled by applying a low-pass filter to the sensitivity numbers \citep{Sigmund_1998}. The convolution operator of the filter is defined as
\begin{align}
\label{eq:convolution_operator}
\begin{split}
	& H_{e'} = r_{\mrm{min}} - \Delta(e,e'), \\
	& \left\{ e'\, \vert\, e' \in \left\{ 1,\dots,\nel \right\}, \Delta(e,e') \leq r_{\mrm{min}} \right\}, \\
	& e \in \left\{ 1,\dots,\nel \right\}
\end{split}
\end{align}
with filter radius $r_{\mrm{min}}$ and center-to-center distance $\Delta(e,e')$ between two elements $e$ and $e'$. We write the smoothed sensitivity values as
\begin{align}
\label{eq:filtered_sensitivities}
	\hat{\alpha}_{e}^{\bullet} = \frac{\sum_{e'=1}^{\nel} \rho_{e'} H_{e'} \alpha_{e'}^{\bullet}}{\rho_{e} \sum_{e'=1}^{\nel} H_{e'}}.
\end{align}
In~\eqref{eq:filtered_sensitivities}, $\bullet$ is a placeholder for $\gamma$ and $\kappa$. To furthermore improve convergence during the iterative process, the sensitivity history of each element is considered \citep{Huang_2007}. Here this is implemented by averaging the sensitivity numbers of the current design iteration $k$ with their counterparts from the previous iteration $k-1$:
\begin{align}
\label{eq:history_stabilization}
	\left[ \hat{\alpha}_{e}^{\bullet} \right]^{k} \leftarrow \frac{1}{2} \left[ \left[ \hat{\alpha}_{e}^{\bullet} \right]^{k} + \left[ \hat{\alpha}_{e}^{\bullet} \right]^{k-1} \right].
\end{align}

\subsection{Optimality criteria design variable update scheme}
\label{sec:update_scheme}
We update the design variables $\gamma_{e}$ and $\kappa_{e}$ by adapting the well-known optimality criteria method \citep{Bendsoe_2004}. The design update of the topological variable $\gamma_{e}$ then reads
\begin{align}
\label{eq:oc_gamma}
	\gamma_{e}^{k+1} =
	\begin{cases}
	 	\max \left( \left[1-\mu\right] \gamma_{e}^{k}, \rhovoid \right) \\ \quad \text{if $\;\gamma_{e}^{k} B_{\gamma}^{k} \leq \max \left( \left[1-\mu \right] \gamma_{e}^{k},\rhovoid \right)$}, \\
	 	\min \left( \left[1+\mu \right] \gamma_{e}^{k}, 1 \right) \\ \quad \text{if $\;\gamma_{e}^{k} B_{\gamma}^{k} \geq \min \left( \left[1+\mu \right] \gamma_{e}^{k},1 \right)$}, \\
	 	\gamma_{e}^{k} B_{\gamma}^{k} \quad \text{otherwise.}
	 \end{cases}
\end{align}
Analogously, the microstructural relative density $\kappa_{e}$ is updated via
\begin{align}
\label{eq:oc_kappa}
	\kappa_{e}^{k+1} =
	\begin{cases}
	 	\max \left( \left[1-\mu\right] \kappa_{e}^{k}, \rhomin \right) \\\quad \text{if $\;\kappa_{e}^{k} B_{\kappa}^{k} \leq \max \left( \left[1-\mu \right] \kappa_{e}^{k},\rhomin \right)$}, \\
	 	\min \left( \left[1+\mu \right] \kappa_{e}^{k}, \rhomax \right) \\\quad \text{if $\;\kappa_{e}^{k} B_{\kappa}^{k} \geq \min \left( \left[1+\mu \right] \kappa_{e}^{k},\rhomax \right)$}, \\
	 	\kappa_{e}^{k} B_{\kappa}^{k} \quad \text{otherwise.}
	 \end{cases}
\end{align}
The change per iteration is limited by the move parameter $\mu$. The scale factors $B_{\gamma}^{k}$ and $B_{\kappa}^{k}$ result from the ratio of the sensitivity numbers to the gradient of the volume constraint~\eqref{eq:relaxed_constraint_volume} in iteration $k$. We write these scale factors as
\begin{align}
	\label{eq:oc_scale_factor}
	B_{\gamma}^{k} = \left[ \frac{ \left[ \hat{\alpha}_{e}^{\gamma} \right]^{k}}{\Lambda^{k} \kappa_{e}^{k} \lvert \Omega_{e} \rvert} \right]^{\eta}, \quad B_{\kappa}^{k} = \left[ \frac{ \left[ \hat{\alpha}_{e}^{\kappa} \right]^{k}}{\Lambda^{k} \gamma_{e}^{k} \lvert \Omega_{e} \rvert} \right]^{\eta}
\end{align}
with $\eta$ and $\Lambda^{k}$ denoting an artificial damping parameter used for convergence control and a Lagrangian multiplier enforcing the inequality constraint on the volume fraction $V_{\mrm{frac}}$, respectively. As demonstrated by Sigmund \citep{Sigmund_2001}, $\Lambda^{k}$ can be obtained by means of a bisection algorithm. 

\subsection{Convergence criterion}
\label{sec:convergence_criterion}
The optimization procedure is carried out until all constraints as well as the convergence criterion
\begin{align}
	\label{eq:convergence_criterion}
	\frac{\sum_{i=1}^{N} \left| c^{k-i+1} - c^{k-N-i+1} \right|}{\sum_{i=1}^{N} c^{k-i+1}} \leq \epsilon \ll 1
\end{align}
are satisfied \citep{Huang_2007}. In this study we chose $N = 5$ such that a (locally) optimal solution is assumed when the compliance is stable over at least 10 successive design iterations.

\section{Data-driven multiscale material modeling of quasi-stochastic lattices}
\label{sec:material_modeling}
Data-driven multiscale material modeling methods are gaining increasing attention due to their flexibility and computational efficiency. In this section, we introduce an isotropic neural network model suitable for topology optimization applications. The training data is generated through computational homogenization. Although our model presented in this section focuses on the deformation behavior of a quasi-stochastic lattice with tetrahedral unit cells, we emphasize that it is a general approach that can model any isotropic cellular material.

\subsection{Computational homogenization framework for beam-lattices}
\label{sec:homogenization}
To capture the multiscale behavior of a lattice, a scale separation between the macroscopic material with spatially varying properties and its microstructure is assumed. The effective, macroscopic counterparts $\bSigma$ and $\bEpsilon$ of the microstructural stress $\bsigma$ and strain $\bepsilon$ can then be computed from averages over the domain $\omega$ of a statistically representative volume element (RVE) \citep{Zohdi_2004}:
\begin{subequations}
\label{eq:effective_quantities}
\begin{align}
	\bEpsilon &= \langle \bepsilon \rangle = \frac{1}{\lvert \omega \rvert} \int_{\omega} \bepsilon \d{\omega}, \label{eq:effective_strain}\\
	\bSigma &= \langle \bsigma \rangle = \frac{1}{\lvert \omega \rvert} \int_{\omega} \bsigma \d{\omega}. \label{eq:effective_stress}
\end{align}
\end{subequations}
Objective of the homogenization procedure now is to find the effective stiffness tensor $\C^{*}$ such that the relation
\begin{align}
	\label{eq:homogenization_objective}
	\bSigma = \C^{*} : \bEpsilon
\end{align}
is fulfilled. As we assume the rods of the lattice to deform like beams, we apply periodic boundary conditions of the form
\begin{align}
	\label{eq:periodic_bcs}
	\bu_{\mrm{mic}}^{+} - \bu_{\mrm{mic}}^{-} = \bEpsilon \left[ \bx^{+} - \bx^{-} \right], \quad \boldsymbol{\theta}^{+} = \boldsymbol{\theta}^{-}
\end{align}
to the RVE to satisfy Hill's energy condition
\begin{align}
	\label{eq:hill_condition}
	\bEpsilon : \bSigma = \frac{1}{\lvert \omega \rvert} \int_{\omega} \bepsilon : \bsigma \d{\omega}.
\end{align}
Here, $\bu_{\mrm{mic}}^{+}$, $\bu_{\mrm{mic}}^{-}$ and $\bx^{+}$, $\bx^{-}$ are displacements and joint node positions on periodically connected boundary faces, respectively, whereas $\boldsymbol{\theta}^{+}$, $\boldsymbol{\theta}^{-}$ indicate rotational degrees of freedom. Since~\eqref{eq:effective_stress} is not trivial to evaluate for discrete structures such as lattices, using the divergence theorem, the volume average over the stress tensor is converted to a boundary integral resulting in
\begin{align}
	\label{eq:stress_boundary_integral}
	\bSigma = \frac{1}{\lvert \omega \rvert} \sum_{n \in \partial \omega} \boldsymbol{f}_{\mrm{mic}}^{n} \otimes \bx^{n}
\end{align}
with internal force vector $\boldsymbol{f}_{\mrm{mic}}^{n}$ at each joint node $n$ located on the boundary faces $\partial \omega$ of the RVE \citep{Gaertner_2021}. Solving six static equlibrium boundary value problems on the RVE with appropriately chosen effective strains
\begin{align}
\begin{split}
	\overline{\bEpsilon}_{1} &= \varepsilon^{*} \begin{bmatrix} 1 & 0 & 0 & 0 & 0 & 0 \end{bmatrix}^{\T}, \\
	\overline{\bEpsilon}_{2} &= \varepsilon^{*} \begin{bmatrix} 0 & 1 & 0 & 0 & 0 & 0 \end{bmatrix}^{\T}, \\
	\overline{\bEpsilon}_{3} &= \varepsilon^{*} \begin{bmatrix} 0 & 0 & 1 & 0 & 0 & 0 \end{bmatrix}^{\T}, \\
	\overline{\bEpsilon}_{4} &= \varepsilon^{*} \begin{bmatrix} 0 & 0 & 0 & 2 & 0 & 0 \end{bmatrix}^{\T}, \\
	\overline{\bEpsilon}_{5} &= \varepsilon^{*} \begin{bmatrix} 0 & 0 & 0 & 0 & 2 & 0 \end{bmatrix}^{\T}, \\
	\overline{\bEpsilon}_{6} &= \varepsilon^{*} \begin{bmatrix} 0 & 0 & 0 & 0 & 0 & 2 \end{bmatrix}^{\T}, \\
	\varepsilon^{*} &\in \R \setminus \left\{ 0 \right\}
\end{split}
\end{align}
as part of the boundary conditions~\eqref{eq:periodic_bcs} finally yields 36 equations from which the 21 independent components of $\C^{*}$ are computed.

\subsection{Modeling of quasi-stochastic representative volume elements}
\label{sec:volume_elements}
\begin{figure}[htb]
	\centering
	\includegraphics[scale=0.95]{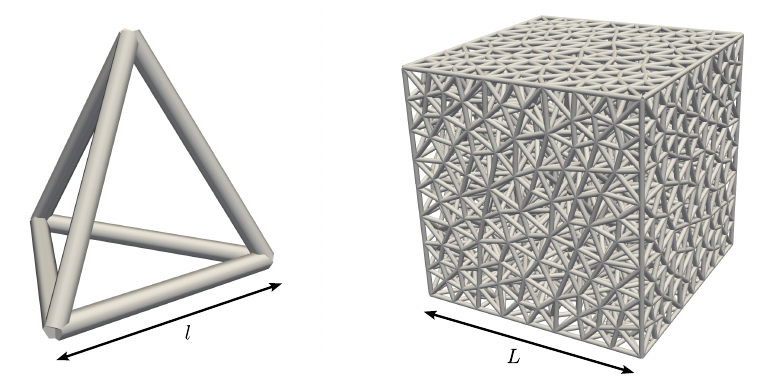}
	\caption{Tetrahedral unit cell and cubic RVE composed of 2,845 unit cells. The aspect ratio is $a = 0.16$.}
	\label{fig:tetrahedral_RVE}
\end{figure}

For lattices made of regular cubic unit cells, usually the unit cell itself is used as the RVE in the homogenization procedure. However, similar to the work by Daynes \citep{Daynes_2023}, our lattice of interest is quasi-stochastic and consists of tetrahedral cells. Therefore, we construct a cubic RVE with size $L \times L \times L$ composed of multiple cells such that the microstructure is represented statistically accurate, see Figure~\ref{fig:tetrahedral_RVE}. The tetrahedral cells are generated from Delaunay triangulation. For simplicity, their edges are replaced by Euler-Bernoulli beam elements with uniform diameter $d$ and circular cross section connected through rigid joints. Other beam theories and assumptions for the joints could be considered as well here \citep{Weeger_2019}.

The algorithm we use constrains the lengths of the edges of each tetrahedral cell to be equal to a uniform strut length $l$. Thus, the RVE can be characterized by the number of struts or cells and the aspect ratio $a = d/l$ as well as Young's modulus $E$ and Poisson's ratio $\nu$ of the base material. Note that the periodic repetition of the RVE in each spatial direction spans an infinite material domain and, hence, the stiffness of overlapping struts is reduced by multiplication with weighting factors $w_{E} \in \left\{ 1/4, 1/2, 1 \right\}$ for edge, face and inner elements, respectively. 

It is well-known that increasing the irregularity of features in a cellular material yields isotropic deformation behavior \citep{Luxner_2007, Zhu_2000, Daynes_2023} which we also expect for the quasi-stochastic RVE. To assess the isotropy, we use the relative anisotropy
\begin{align}
	\label{eq:relative_elastic_anisotropy}
	\Delta_{\mrm{iso}} = \frac{\lVert \C^{*} - \C_{\mrm{iso}}^{*} \rVert}{\lVert \C^{*} \rVert}
\end{align}
proposed by Gatt et al. \citep{Gatt_2005} as a measure. Thereby, $\C_{\mrm{iso}}$ is the isotropic tensor closest to $\C^{*}$ in the Euclidean distance
\begin{align}
	\label{eq:isotropic_approximation}
	\overline{\C}^{*}_{\mrm{iso}} = \mrm{tr} \left( \boldsymbol{Q} \, \overline{\C}^{*} \, \boldsymbol{Q} \boldsymbol{A}_{1} \right) \boldsymbol{A}_{1} + \mrm{tr} \left( \boldsymbol{Q} \, \overline{\C}^{*} \boldsymbol{Q} \boldsymbol{A}_{2} \right) \boldsymbol{A}_{2}
\end{align}
with the overline denoting Voigt notation again \citep{Gazis_1963}. The definitions of the constant projection matrices $\boldsymbol{Q}$, $\boldsymbol{A}_{1}$ and $\boldsymbol{A}_{2}$ are
\begin{align}
\begin{gathered}
	\boldsymbol{Q} = \begin{bmatrix}
		1 & 0 & 0 & 0 & 0 & 0 \\
		0 & 1 & 0 & 0 & 0 & 0 \\
		0 & 0 & 1 & 0 & 0 & 0 \\
		0 & 0 & 0 & 2 & 0 & 0 \\
		0 & 0 & 0 & 0 & 2 & 0 \\
		0 & 0 & 0 & 0 & 0 & 2 \\
	\end{bmatrix},\quad
	\boldsymbol{A}_{1} = \frac{1}{3} \begin{bmatrix}
		1 & 1 & 1 & 0 & 0 & 0 \\
		1 & 1 & 1 & 0 & 0 & 0 \\
		1 & 1 & 1 & 0 & 0 & 0 \\
		0 & 0 & 0 & 0 & 0 & 0 \\
		0 & 0 & 0 & 0 & 0 & 0 \\
		0 & 0 & 0 & 0 & 0 & 0
	\end{bmatrix},\\
	\boldsymbol{A}_{2} = \frac{1}{6 \sqrt{5}} \begin{bmatrix}
		4 & -2 & -2 & 0 & 0 & 0 \\
		-2 & 4 & -2 & 0 & 0 & 0 \\
		-2 & -2 & 4 & 0 & 0 & 0 \\
		0 & 0 & 0 & 3 & 0 & 0 \\
		0 & 0 & 0 & 0 & 3 & 0 \\
		0 & 0 & 0 & 0 & 0 & 3
	\end{bmatrix}.
\end{gathered}
\end{align}

Based on~\eqref{eq:relative_elastic_anisotropy} and~\eqref{eq:isotropic_approximation} we conclude that the RVE from Figure~\ref{fig:tetrahedral_RVE} consisting of 2,845 tetrahedral cells and 4,137 struts can be characterized isotropic since we obtain $\Delta_{\mrm{iso}} = 4.3\,\%$ already for a very small aspect ratio $a = 0.01$. By increasing $a$, the relative anisotropy is reduced further. Therefore, in the rest of our study we use this RVE topology with variable aspect ratio and base material in our homogenization framework. Furthermore, due to the near-isotropic behavior obtained for $\C^{*}$, it is reasonable to proceed with its isotropic projection $\C^{*}_{\mrm{iso}}$ as the effective stiffness of the lattice material, i.e. we replace
\begin{align}
\label{eq:replace_effective_stiffness}
	\C^{*} \leftarrow \C^{*}_{\mrm{iso}}.
\end{align}

\subsection{Estimation of the microstructural relative density}
\label{sec:relative_density}
\begin{figure}[htb]
	\centering
	\includegraphics[scale=0.95]{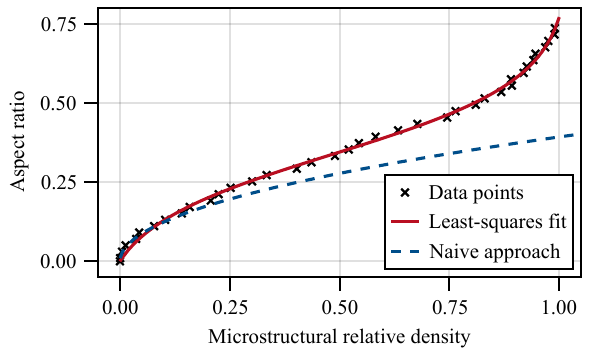}
	\caption{Functional relation between the microstructural relative density $\kappa$ and the aspect ratio $a$ computed through least-squares fitting. The outcome of the naive approach deviates significantly from the data obtained through Monte Carlo integration.}
	\label{fig:density_aspect_mapping}
\end{figure}

Since the RVE is controlled through the aspect ratio as well as the uniform stiffness parameters of the struts, we conclude the effective lattice stiffness to be an explicit function $\C^{*} = \C^{*} \!\left( \bp \right)$ with $\bp = \bp \left( \kappa, E, \nu \right) = \begin{bmatrix} a(\kappa) & E & \nu \end{bmatrix}^{\T}$. However, the mapping $a = a \left( \kappa \right)$ is still unknown. To model this relation, we need estimates of $\kappa$ for different values $a = \hat{a}$. For that, it seems natural to sum up the volumes of all individual struts being part of the corresponding RVE and to divide the result by the total volume $L^{3}$ of the cube afterwards. However, since intersections at the joints are disregarded by this naive method, it is only useful for low aspect ratios. Therefore, we apply a Monte Carlo integration algorithm \citep{Souza_2018} that iteratively generates random points in the RVE. The microstructural relative density might then be estimated from
\begin{align}
	\label{eq:monte_carlo_integration}
	\kappa^{i}\left( \hat{a} \right) \approx \frac{\widetilde{m}^{i} \left( \hat{a} \right)}{m^{i}}
\end{align}
with the total number of points $m^{i}$ and $\widetilde{m}^{i} \left( \hat{a} \right)$ being the number of points that intersect with a lattice strut in integration step $i$. The generation of new points is aborted as soon as $\kappa^{i} \left( \hat{a} \right)$ changes less than 0.1\,\% compared to the previous iteration $i-1$. All data points obtained from this algorithm for different $\hat{a}$ are visualized in Figure~\ref{fig:density_aspect_mapping}. 

Based on the arrangement of the points and without loss of generality, we propose a parametrized version of the inverse Sigmoid function
\begin{equation}
\label{eq:inverse_sigmoid}
\begin{aligned}
	a \left( \kappa \right) = c_{1} &\ln \left( \frac{1}{c_{2} c_{3}} - 1 \right) - c_{1} \ln \left( \frac{1}{c_{2} \left( \kappa + c_{3} \right)} - 1 \right)
\end{aligned}
\end{equation}
to connect the aspect ratio with the microstructural density. We note that the first term in~\eqref{eq:inverse_sigmoid} ensures that $a \left( \kappa = 0 \right) = 0$. The values for the parameters $c_{i}$ have been found via least-squares fitting to the data points:
\begin{align}
	\label{eq:aspect_density_constants}
	c_{1} = 0.11882, \quad c_{2} = 0.91991, \quad c_{3} = 0.05956.
\end{align}

\subsection{Parametric material model}
\label{sec:parametric_model}
We model the isotropic lattice stiffness tensor $\C^{*} = \C^{*} \left( \bp \right)$ using a physics-augmented neural network. Our isotropic model is thereby an extension of the model proposed by Shojaee et al. \citep{Shojaee_2023} that was originally designed for rhombic materials.

\subsubsection{Isotropic Cholesky matrices}
We know that the strain energy density requires the following to be satisfied:
\begin{align}
	\Psi = \frac{1}{2} \bEpsilon : \C^{*} : \bEpsilon > 0, \quad \forall \bEpsilon \neq \boldsymbol{0},
\end{align}
as a necessary physical constraint. As a consequence, the stiffness tensor must be positive definite. Adopting Voigt notation, we can enforce positive definiteness by the Cholesky decomposition 
\begin{align}
	\label{eq:cholesky_decomposition}
	\overline{\C}^{*} = \bG \bG^{\T}
\end{align}
where $\bG = \bG \left( \bp \right)$ is a lower triangular matrix with positive diagonal elements \citep{Haddad_2009}. For isotropic material symmetry, $\bG$ consists of six generally unequal but not independent coefficients:
\begin{align}
	\label{eq:cholesky_matrix}
	\bG = \begin{bmatrix}
		G_{11} & 0 & 0 & 0 & 0 & 0 \\
		G_{21} & G_{22} & 0 & 0 & 0 & 0 \\
		G_{21} & G_{32} & G_{33} & 0 & 0 & 0 \\
		0 & 0 & 0 & G_{44} & 0 & 0 \\
		0 & 0 & 0 & 0 & G_{44} & 0 \\
		0 & 0 & 0 & 0 & 0 & G_{44}
	\end{bmatrix}.
\end{align}
Choosing $G_{11}$ and $G_{44}$ to be the independent isotropy coefficients, the dependent matrix components can be identified through standard algebraic operations with the following result:
\begin{subequations}
\label{eq:cholesky_coefficients}
\begin{align}
	G_{22} &= 2 G_{44} \sqrt{1 - \left[ \frac{G_{44}}{G_{11}} \right]^{2}}, \\
	G_{33} &= G_{44} \sqrt{\frac{3 - 4 \left[ G_{44} / G_{11} \right]^{2}}{1 - \left[ G_{44} / G_{11} \right]^{2}}}, \\
	G_{21} &= G_{11} - 2 \left[ \frac{G_{44}}{G_{11}} \right]^{2}, \\
	G_{32} &= G_{44} \frac{1 - 2 \left[ G_{44} / G_{11} \right]^{2}}{\sqrt{1 - \left[ G_{44} / G_{11} \right]^{2}}}
\end{align}
\end{subequations}
Positivity of all diagonal entries of $\bG$ is ensured through satisfaction of the constraint
\begin{align}
	\label{eq:positivity_constraint}
	G_{11} > \frac{2}{\sqrt{3}} G_{44} > 0
\end{align}
that is derived from the requirement $G_{33} > 0$. For details on the derivation of the isotropic Cholesky matrix and its components, we refer the interested reader to~\ref{sec:appendix_cholesky}.

\subsubsection{Neural network approximation}
\begin{figure}[htb]
	\centering
	\includegraphics[scale=0.95]{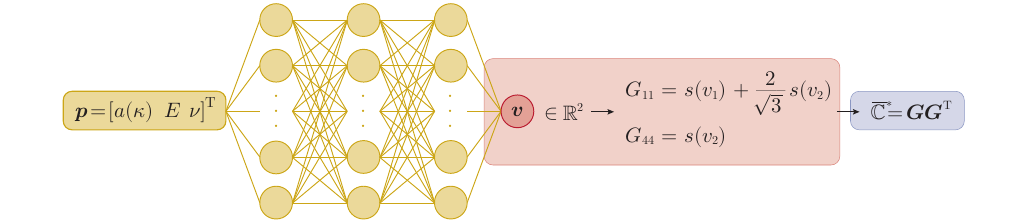}
	\caption{Schematic design of the physics-augmented neural network material model adapted from \cite{Shojaee_2023}.}
	\label{fig:neural_network}
\end{figure}

In this work, instead of finding an explicit form of $\bG \left( \bp \right)$, we utilize a feed forward neural network (FFNN) with three hidden layers and 64 nodes per layer, implemented through the Julia Flux framework \citep{Innes_2018}. Analogously to \cite{Shojaee_2023}, we choose the activation function of each layer to be the softplus function $s(x) = \ln \left( 1 + \exp \left( x \right) \right)$ which acts component-wise on non-scalar inputs and always yields non-negative outputs. Thus, the map between $\bp$ and the FFNN output vector $\bv = \begin{bmatrix} v_{1} & v_{2} \end{bmatrix}^{\T}$ is
\begin{equation}
\label{eq:neural_network}
\begin{aligned}
	\bv = \bW^{4} &s \left( \bW^{3} s \left( \bW^{2} s \left( \bW^{1} \bp + \bb^{1} \right) + \bb^{2} \right) + \bb^{3} \right) + \bb^{4}
\end{aligned}
\end{equation}
with weight matrices
\begin{equation}
\label{eq:weight_matrices}
\begin{gathered}
	\bW^{1} \in \R^{64 \times 3},\quad \bW^{2},\; \bW^{3} \in \R^{64 \times 64}, \quad \bW^{4} \in \R^{2 \times 64}, 	
\end{gathered}
\end{equation}
and bias vectors
\begin{align}
	\label{eq:bias_vectors}
	\bb^{1},\; \bb^{2},\; \bb^{3} \in \R^{64},\quad \bb^{4} \in \R^{2}.
\end{align}
The Cholesky components of interest finally result from $\bv$ as
\begin{subequations}
\begin{align}
	\label{eq:neural_network_cholesky_coefficients}
	G_{11} &= s \left( v_{1} \right) + \frac{2}{\sqrt{3}} s \left( v_{2} \right) \\
	G_{44} &= s \left( v_{2} \right)
\end{align}
\end{subequations}
such that~\eqref{eq:positivity_constraint} is satisfied. The overall architecture of the FFNN is illustrated in Figure~\ref{fig:neural_network}. 

Note, that a neural network modeling two independent and isotropic material parameters such as the effective Lam{\'e} constants instead of the Cholesky components can yield feasible stiffness tensors in a similar way. In that case, material symmetry is enforced directly by construction of $\C^{*}$. However, the constraints on the neural network output have to be modified appropriately to the material parameters of interest.

\subsubsection{Training of the neural network}
To find values for the hyperparameters $\bW^{i}$ and $\bb^{i}$ that approximate $\bG \left( \bp \right)$ best, the dataset 
\begin{align}
	\label{eq:training_dataset}
	\mathcal{D}_{\mrm{t}} = \left\{ \left( \hat{\bp}^{1}, \hat{\bG}^{1} \right), \dots, \left( \hat{\bp}^{n_{\mrm{t}}}, \hat{\bG}^{n_{\mrm{t}}} \right) \right\}
\end{align}
is introduced. The Cholesky matrices $\hat{\bG}^{i}$ are obtained from effective stiffness tensors computed via the homogenization and isotropic projection framework presented in Sections~\ref{sec:homogenization} and~\ref{sec:volume_elements} for each vector of parameters $\hat{\bp}^{i}$. Next, the nonlinear AMSGrad optimizer with both decay parameters set to 0.9 minimizes the mean-squared error (MSE) of the deviation of the model predictions from the training data
\begin{align}
	\label{eq:mean_squared_error}
	\text{MSE} = \frac{1}{2 n_{\mrm{t}}} \sum_{i = 1}^{n_{\mrm{t}}} \sum_{j \in \left\{ 1, 4 \right\}} \left[ \hat{G}_{jj}^{i} - G_{jj} \left( \hat{\bp}^{i} \right) \right]^{2}.
\end{align}
After $250 \times 10^{3}$ training iterations (epochs), the loss on $\mathcal{D}_{\mrm{t}}$ as well as on an unseen validation dataset $\mathcal{D}_{\mrm{v}}$ has decreased from $\text{MSE} \approx 130$ to $\text{MSE} \approx 0.001$ such that the model is assumed to represent $\bG \left( \bp \right)$ with satisfactory precision.

\subsubsection{Model validation} 
\begin{figure}[htb]
	\centering
	\includegraphics[scale=0.95]{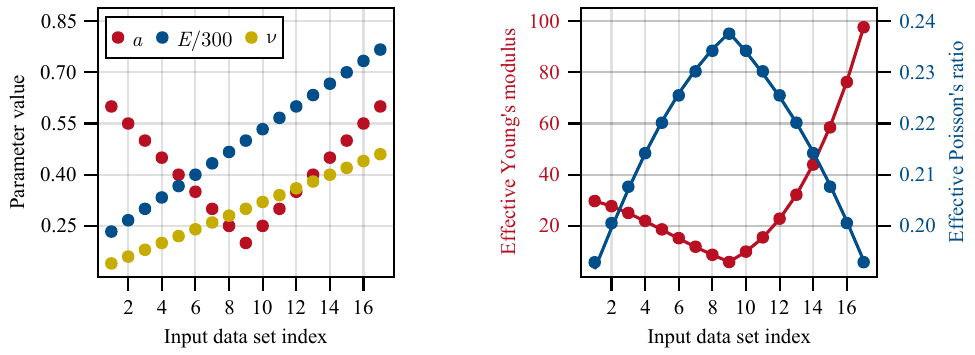}
	\caption{Accuracy of the FFNN on the continuous test data set $\mathcal{D}_{\mrm{c}}$. The parameter triplets used as model inputs are given on the left side. The effective Young's modulus $E^{*}$ and Poisson's ratio $\nu^{*}$ in the right graph are obtained from $\C^{*}$ and~\eqref{eq:effective_material_parameters}. In both graphs, solid dots denote data points and continuous lines are the predictions of the neural network model.}
	\label{fig:model_evaluation}
\end{figure}

To further investigate how well the model actually performs on unseen data, we consider a data set $\mathcal{D}_{\mrm{c}}$ of 17 continuously varying parameter triplets and their corresponding stiffness tensors. Based on the results shown in Figure~\ref{fig:model_evaluation}, we conclude that the FFNN model is able to predict the effective Young's modulus and Possion's ratio
\begin{subequations}
\label{eq:effective_material_parameters}
\begin{align}
	E^{*} &= \frac{\overline{\C}^{*}_{66} \left[ 3 \, \overline{\C}^{*}_{12} + 2 \, \overline{\C}^{*}_{66} \right]}{\overline{\C}^{*}_{12} + \overline{\C}^{*}_{66}}, \label{eq:effective_youngs_modulus} \\
	\nu^{*} &= \frac{\overline{\C}^{*}_{12}}{2 \left[ \overline{\C}^{*}_{12} + \overline{\C}^{*}_{66} \right]} \label{eq:effective_poissons_ratio}
\end{align}
\end{subequations}
and, therefore, also $\C^{*}$ reliably. We observe that $\nu^{*}$ seems to be influenced mainly by the aspect ratio $a$, while $E^{*}$ heavily depends on the aspect ratio as well as the Young's modulus of the base material $E$.

\section{Numerical examples, results and discussion}
\label{sec:results}
In this section, we consider three example problems that are suitable to demonstrate and validate our multiscale topology optimization framework. 

\subsection{Messerschmitt-B\"olkow-Blohm beam}
\label{sec:mbb}
\begin{figure}[htb]
	\centering
	\includegraphics[scale=0.95]{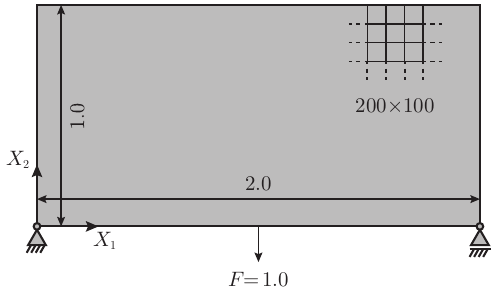}
	\caption{Setup of the MBB beam example.}
	\label{fig:mbb}
\end{figure}

The two-dimensional Messerschmitt-B\"olkow-Blohm (MBB) beam problem is widely used as a benchmark in the field of topology optimization. The general setup of this example is illustrated in Figure~\ref{fig:mbb}. The length and height of the beam are set to 2.0 and 1.0, respectively, and the domain is discretized with a mesh consisting of $200 \times 100$ bi-linear, quadrilateral Lagrangian finite elements. An external force with magnitude $F = 1.0$ is applied to the center point of the bottom edge of the structure.

The effective material behavior is described by the plane strain formulation of the parametric model introduced in Section~\ref{sec:parametric_model}. For that, the struts of the lattice have uniform and constant Young's modulus $E = 210.0$ and Poisson's ratio $\nu = 0.3$. Furthermore, all relative density values between the threshold $\rhomin = 0.1$ and $\rhomax = 0.25$ are assumed to be additively manufacturable. The fabrication of the optimized structure shall consume at most 15\,\% of the material a solid structure requires, i.e. we set $V_{\mrm{frac}} = 0.15$. In addtion, we assume an initial relative density distribution that already satisfies this constraint by choosing $\gamma_{e}^{0} = 1.0$ and $\kappa_{e}^{0} = 0.15$ for all elements $e$ in the domain. Intermediate values for the topological design variables in the constitutive material tensor~\eqref{eq:penalized_material_tensor} are penalized using $p = 3$. Moreover, we apply a sensitivity filter with $r_{\mrm{min}} = 0.06$. The damping and move parameters used to update the design variables are $\eta = 0.5$ and $\mu = 0.05$, respectively. Convergence is evaluated from~\eqref{eq:convergence_criterion} with $\epsilon = 10^{-4}$.

\begin{figure}[htb]
	\centering
	\includegraphics[scale=0.95]{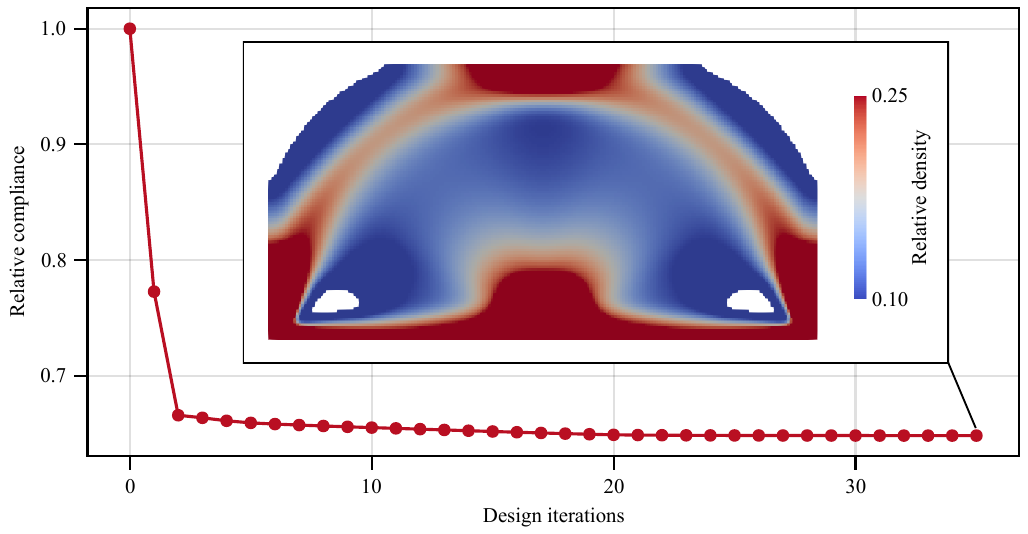}\vspace*{-5mm}
	\caption{Convergence history with the optimized topology and relative density grading for the MBB beam example after 35 iterations. The compliance in the convergence plot is scaled relatively to the compliance of the initial uniform density structure. In the contour plot all finite elements with a relative density lower than the threshold value $\rhomin = 0.1$ are not displayed.}
	\label{fig:convergence_mbb}
\end{figure}

Figure~\ref{fig:convergence_mbb} shows a compliance convergence plot as well as the topology and density distribution of the final structure. It can be seen that after only two design variable update iterations the overall compliance of the structure has already decreased by over 33\,\%. After that, the compliance constantly decreases further until the convergence criterion is satisfied after 35 iterations. The converged design is found to be 35.2\,\% stiffer than the initial uniform density structure although it consumes the same amount of base material. Constraint~\eqref{eq:relaxed_constraint_volume} is active for the whole optimization process.

Considering the optimized MBB beam, we observe an area made up of material with low relative density supporting the stiffness in the center of the structure which is enclosed by higher relative density material. The material with the maximum possible relative density accumulates primarily around the support regions and the point of force transmission. From that point, material with intermediate density furthermore spreads in the diagonal directions. In the upper corners material is removed completely. Our results coincide qualitatively with results from similar studies that can be found in the literature, see for example \cite{Gangwar_2021} and \cite{Lu_2023}.

\begin{figure}[htb]
	\centering
	\includegraphics[scale=0.95]{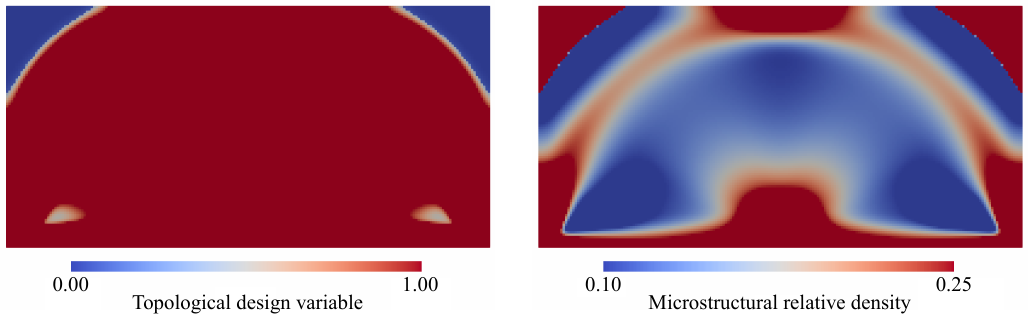}\vspace*{-5mm}
	\caption{Optimized distribution of the topological design variable field $\gamma$ (left) and the microstructural relative density field $\kappa$ (right). The multiplicative superposition of both fields yields the relative density field $\rho$. Note that the colorbars are scaled differently.}
	\label{fig:gamma_kappa_mbb}
\end{figure}

The already examined relative density field results from the multiplicative superposition of the topological field $\gamma$ and the microstructural relative density field $\kappa$, both visualized in Figure~\ref{fig:gamma_kappa_mbb}. The penalization approach~\eqref{eq:penalized_material_tensor} turns out to be effective in pushing $\gamma$ towards its limit values such that a 0-1 distribution is obtained. Only in the very small transition areas intermediate values for $\gamma$ are present. In the material domain, i.e. where $\gamma = 1$, it is trivial that the microstructural relative density $\kappa$ coinicides with the macroscale density $\rho$. Furthermore, even if the effects on the compliance of the structure are negligible due to the penalization of the effective stiffness tensor, the microstructural density takes its upper limit value in the void domain. The reason is that high values for $\kappa$ in this area practically do not contribute to the total volume of the structure.

\subsection{Cantilever beam}
\label{sec:cantilever_beam}
\begin{figure}[htb]
	\centering
	\includegraphics[scale=0.95]{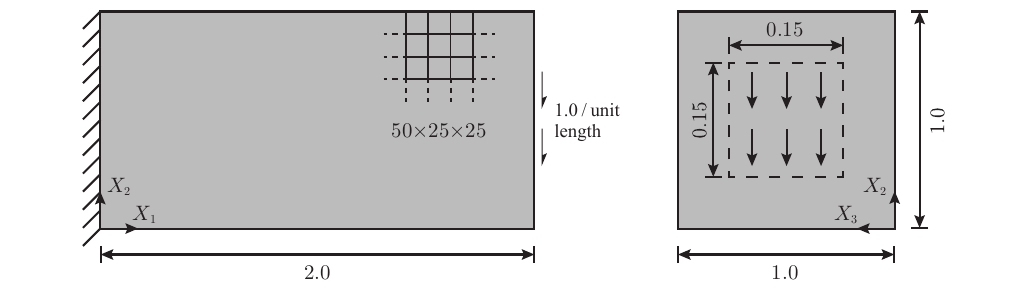}\vspace*{-5mm}
	\caption{Setup of the three-dimensional cantilever beam example.}
	\label{fig:cantilever}
\end{figure}

The second example we want to test our optimization framework on is the three-dimensional cantilever beam sketched in Figure~\ref{fig:cantilever}. The length, width and height of the beam are 2.0, 1.0 and 1.0, respectively. The left end is fixed and a traction force of magnitude 1.0 per unit length is applied in negative $X_{2}$-direction to a region of size $0.15 \times 0.15$ located at the center of the right square surface. The mesh used to discretize the design space consists of $50 \times 25 \times 25$ tri-linear, hexahedral finite elements with Lagrangian shape functions. The base material parameters are $E = 210.0$ and $\nu = 0.3$ again. We impose constraints on the manufacturability and the available material with $\kappa_{e} \in \left[ 0.05, 0.4 \right]$ and $V_{\mrm{frac}} = 0.1$. To test the algorithm with an infeasible initial guess, constraint~\eqref{eq:relaxed_constraint_volume} is violated using $\gamma_{e}^{0} = 1.0$ and $\kappa_{e}^{0} = 0.4$. The remaining parameters regarding the stability and convergence of the algorithm are chosen to be the same as in Section~\ref{sec:mbb}.

\begin{figure}[htb]
	\centering
	\includegraphics[scale=0.95]{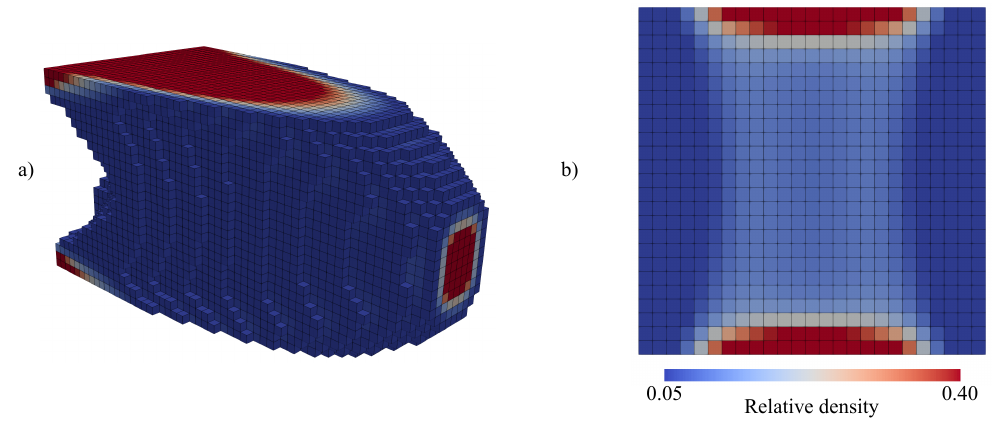}\vspace*{-2mm}
	\caption{a) Optimized topology and density grading of the cantilever beam. b) Front view with a cut through the $X_{2}$-$X_{3}$ plane at $X_{1} = 1.0$. Finite elements representing void are not displayed.}
	\label{fig:cantilever_optimization}
\end{figure}

The algorithm converges after 301 iterations with compliance $c = 2.89$. It results in a structure that is over 74.1\,\% stiffer than a simple lattice cantilever satisfying the constraints with a uniform relative density $\rho = 0.1$. The volume constraint becomes and stays active after seven design variable updates.

Figure~\ref{fig:cantilever_optimization} shows the final design of the cantilever beam. Similar to the MBB beam example, material with high relative density gathers around highly stressed regions such as the upper and lower surface that undergo the most tension and compression, respectively, as well as the traction surface. The rest of the outer shell of the beam consists mainly of low-density material. Cutting through the beam reveals that material with intermediate relative density increases the stiffness in the bulk volume.

\subsection{Jet engine bracket}
\label{sec:jet_engine_bracket}
\begin{figure}[htb]
	\centering
	\includegraphics[scale=0.95]{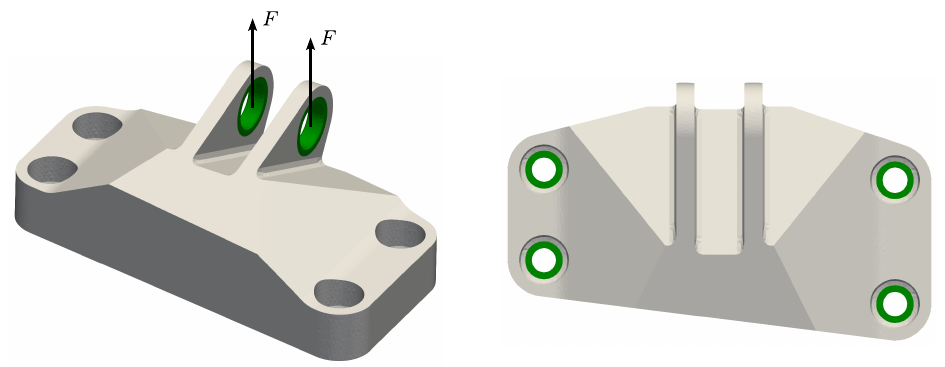}
	\caption{Setup of the aircraft jet engine bracket example.}
	\label{fig:bracket}
\end{figure}

To demonstrate the applicability of our method to real-world engineering problems, we consider the jet engine bracket given in Figure~\ref{fig:bracket}. This aircraft component was introduced in 2013 as part of a community design challenge \citep{GrabCAD_2013} and has since then been used by researchers to test and validate their topology optimization algorithms, see e.g. \cite{Tyflopoulos_2021}. 

In Figure~\ref{fig:bracket}, the regions colored in green serve as support and load application zones and remain unchanged in the optimization procedure. They are made of solid, isotropic and linear elastic material, which is the same as the lattice base material such that both can be defined by the stiffness parameters $E = 210.0$ and $\nu = 0.3$. Fixed supports are applied to the smaller bolt holes and two effective forces $F = 10.0$ act in vertical direction, which are transferred to corresponding surface tractions on the upper half surface of each of the larger pin holes. 

\begin{figure}[htb]
	\centering
	\includegraphics[scale=0.95]{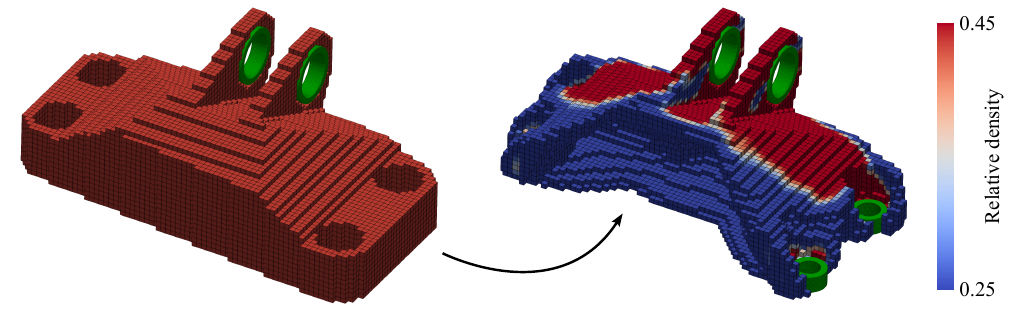}\vspace*{-5mm}
	\caption{Initial voxel mesh with uniform density distribution (left) and optimized topology and density grading of the jet engine bracket (right). Elements with a relative density lower than $\rhomin = 0.25$ are interpreted as void and are therefore not displayed.}
	\label{fig:bracket_optimization}
\end{figure}

The computation is performed on a voxel representation of the bracket consisting of 34,610 tri-linear, 8-node hexahedral elements with edge length 2.38 as shown in Figure~\ref{fig:bracket_optimization}. We set the radius of the sensitivity filter $r_{\mrm{min}} = 6.0$ and the remaining algorithmic parameters $\eta = 0.05$, $\mu = 0.01$, $p = 3$ and $\epsilon = 10^{-4}$. The manufacturability of the structure is assumed to be constrained by $\rhomin = 0.25$ as the threshold value and $\rhomax = 0.45$ as the upper limit for the relative density. In addition, the maximum material volume fraction is $V_{\mrm{frac}} = 0.25$. Since we choose a uniform lattice material distribution with $\gamma_{e}^{0} = 1.0$ and $\kappa_{e}^{0} = 0.45$, the inequality constraint~\eqref{eq:relaxed_constraint_volume} is again violated initially.

After 147 design variable updates our optimization procedure converges. The compliance of the jet engine bracket is thereby over 36.9\,\% lower than the compliance of a naive design satisfying the manufacturing constraints with a uniformly distributed relative density equal to the maximum volume fraction $V_{\mrm{frac}}$.

\begin{figure}[htb]
	\centering
	\includegraphics[scale=0.95]{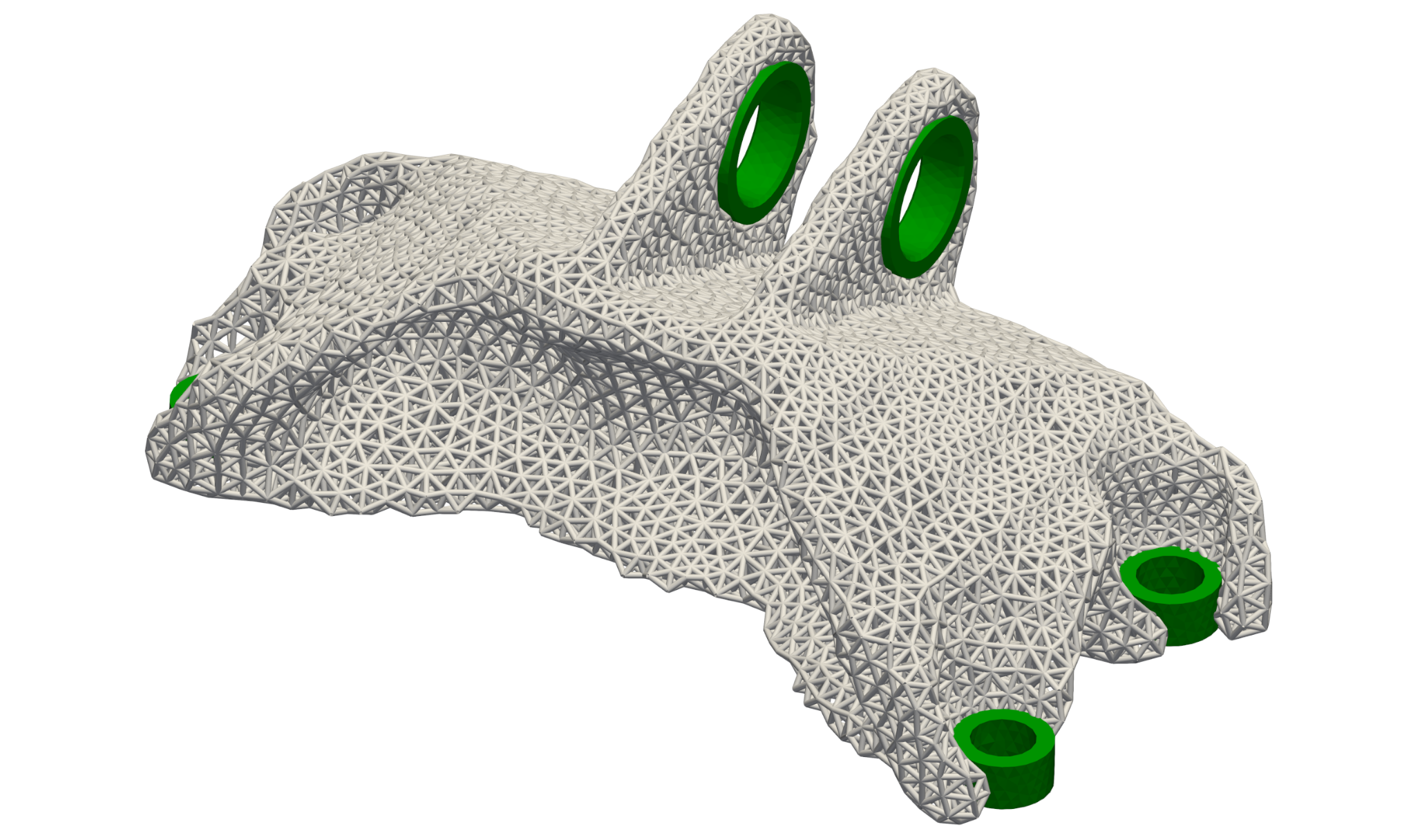}\vspace*{-5mm}
	\caption{Lattice representation of the optimized jet engine bracket.}
	\label{fig:bracket_lattice}
\end{figure}

Figure~\ref{fig:bracket_optimization} illustrates the optimized topology of the structure as well as the relative density grading after convergence. The algorithm mainly drives the generation of a cavity in the bulk volume of the bracket. Further material is removed around the support holes. Elements supporting the flow of forces from the traction boundary to the supports naturally gather most of the available material to increase the stiffness. In contrast to the examples presented in Sections~\ref{sec:mbb} and~\ref{sec:cantilever_beam}, only very few elements with intermediate relative density are part of the final design. To generate the printable lattice representation of the jet engine bracket from Figure~\ref{fig:bracket_lattice}, we compute a smoothed isosurface that separates the void domain from the material domain by means of the marching cubes algorithm \citep{Wenger_2013}. The enclosed volume is then triangulated under consideration of local cell size contraints
\begin{align}
	l = \frac{d}{a \left( \kappa \right)}
\end{align}
and the assumption of a uniform strut diameter $d = 1.0$.

\section{Conclusion and outlook}
\label{sec:conclusion}
In this article, we present a novel general framework for the simultaneous optimization of both the topology and the relative density grading of cellular structures and materials. The relative density is integrated as a material parameter into an isotropic and physics-augmented neural network material model trained on data obtained from computational homogenization. Although we developed our model for strut-based lattice materials, it can be easily extended to other cellular materials with isotropic deformation behavior and could also be generalized to anisotropic microstructures. Using our method, we successfully achieved stiffness-optimized designs for various two- and three-dimensional engineering structures, considering manufacturing and maximum volume constraints.

Our framework employs a sequential multiscale approach sufficient for optimizing the macroscale density distribution of lattices. In the future, we therefore plan to extend it to a concurrent multiscale method similar to the one presented in \cite{Gangwar_2021}. Leveraging the flexibility of data-driven material modeling, this extension will allow us to consider microscale design variables, enabling the optimization of the unit cell topology at each material point. To ensure connectivity and manufacturability of the unit cells, additional constraints will need to be implemented. Furthermore, the parameters of the base material could be included as design variables.

\appendix
\section{Derivation of isotropic Cholesky coefficients}
\label{sec:appendix_cholesky}
We start with the general Cholesky matrix of size $6 \times 6$ that reads \citep{Haddad_2009}
\begin{align}
	\label{eq:general_cholesky_matrix}
	\bG = \begin{bmatrix}
		G_{11} & 0 & 0 & 0 & 0 & 0 \\
		G_{21} & G_{22} & 0 & 0 & 0 & 0 \\
		G_{31} & G_{32} & G_{33} & 0 & 0 & 0 \\
		G_{41} & G_{42} & G_{43} & G_{44} & 0 & 0 \\
		G_{51} & G_{52} & G_{53} & G_{54} & G_{55} & 0 \\
		G_{61} & G_{62} & G_{63} & G_{64} & G_{65} & G_{66}
	\end{bmatrix}.
\end{align}
For the diagonal elements of $\bG$ it must hold
\begin{align}
	\label{eq:cholesky_diagonals}
	G_{ii} > 0, \quad \forall i \in \left\{ 1, 2, 3, 4, 5, 6 \right\}.
\end{align}
We furthermore know that the stiffness tensor of linear elastic and isotropic materials can be written as a positive definite matrix of the form
\begin{align}
	\label{eq:linear_elasticity}
	\overline{\C}^{*} = \begin{bmatrix}
		C_{11} & C_{12} & C_{12} & 0 & 0 & 0 \\
		C_{12} & C_{11} & C_{12} & 0 & 0 & 0 \\
		C_{12} & C_{12} & C_{11} & 0 & 0 & 0 \\
		0 & 0 & 0 & C_{44} & 0 & 0 \\
		0 & 0 & 0 & 0 & C_{44} & 0 \\
		0 & 0 & 0 & 0 & 0 & C_{44}
	\end{bmatrix}
\end{align}
with the secondary condition \citep{Irgens_2008}
\begin{align}
	\label{eq:secondary_condition}
	C_{11} = C_{12} + 2 C_{44}.
\end{align}
After evaluation of $\overline{\C}^{*} = \bG \bG^{\T}$, one can now equate the coefficients of the matrices which yields a system of equations. At this point, the set $\mathcal{Z}$ of all zero-valued entries of $\bG$ is easily determined under consideration of~\eqref{eq:cholesky_diagonals}:
\begin{align}
	\label{eq:cholesky_zero_set}
	\mathcal{Z} = \left\{ G_{54}, G_{64} \right\} \cup \left\{ G_{ij}\, \vert\, i \in \left\{4, 5, 6 \right\}, j \in \left\{ 1, 2, 3 \right\} \right\}.
\end{align}
The remaining equations available to compute the non-zero Cholesky coefficients are
\begin{align}
\label{eq:system_of_equations}
\begin{split}
	G_{11}^{2} &= G_{21}^{2} + G_{22}^{2}, \\
	G_{11}^{2} &= G_{31}^{2} + G_{32}^{2} + G_{33}^{2}, \\
	G_{21} G_{11} &= G_{31} G_{11}, \\
	G_{21} G_{11} &= G_{31} G_{21} + G_{32} G_{22}, \\
	G_{44} &= G_{55}, \\
	G_{44} &= G_{66}, \\
	G_{11}^{2} &= G_{21} G_{11} + 2 G_{44}.
\end{split}
\end{align}
Since isotropic material behavior can be modeled by only two independent parameters, see~\eqref{eq:secondary_condition}, we assume that $G_{11}$ and $G_{44}$ are arbitrary. The seven unkown and dependent coefficients of $\bG$ are then obtained by solving system~\eqref{eq:system_of_equations}.

\section*{Replication of results}
The results can be reproduced based on the information presented in the manuscript only. All data and the developed Julia code can be made available upon request from the corresponding author.

\section*{CRediT authorship contribution statement}
\textbf{Jonathan Stollberg:} Conceptualization, Formal analysis, Investigation, Methodology, Software, Visualization, Writing -- original draft. \textbf{Tarun Gangwar:} Conceptualization, Methodology, Supervision, Writing -- review and editing.
\textbf{Oliver Weeger:} Conceptualization, Methodology, Supervision, Writing -- review and editing. \textbf{Dominik Schillinger:} Conceptualization, Funding acquisition, Project administration, Supervision, Writing -- review and editing.

\section*{Declaration of competing interest}
The authors declare that they have no conflict of interest.

\section*{Funding}
This work has received funding from the European Research Council (ERC) under the European Union’s Horizon 2020 research and innovation programme (Grant agreement No. 759001). T. Gangwar gratefully acknowledges the support from Anusandhan National Research Foundation (ANRF) through Start-up Grant (Grant No. SRG/2023/000144).

\bibliographystyle{elsarticle-num}
\bibliography{bibliography}
\end{document}